\documentclass[aps,prb,showpacs]{revtex4}
\usepackage{graphicx}

\begin{document}

\title{Oscillations of the Electric-Dipole Echo in Glasses in a
Magnetic Field} 

\author{D. A. Parshin}
\affiliation{Department of Physics, St.Petersburg 
State Polytechnical University, 195251, Polytekhnicheskaya 29, 
St.Petersburg, Russia}

\begin{abstract}
Using a simple diagram technique we derive the electric-dipole echo
amplitude from two-level systems with a quadrupole nuclear
moment in glasses in an external magnetic field. We show, that due to
the quadrupole moment interaction of a tunneling particle with a
gradient of an internal electric field, the echo amplitude experiences
oscillations in rather weak magnetic fields. With an 
increase of the magnetic field, when the Zeeman energy becomes larger
than the quadrupole energy splitting, the average echo amplitude increases
and saturates (for high magnetic fields) at some level which is above
the average level of echo oscillations for small magnetic fields.  
\end{abstract}


\maketitle

\section{INTRODUCTION}
Recently it was experimentally observed that the electric-dipole echo
amplitude in non-magnetic glasses oscillates as a function of applied
magnetic 
field~\cite{LESH,LNHE}.  Similar behavior was found in insulating
crystals with tunneling impurities~\cite{EL}. 
In a recent paper~\cite{WFE} such behavior was attributed to 
quadrupole nuclear moments of tunneling particles interacting with the
magnetic field and with the gradient of an internal electric field.
The purpose of this paper is to formulate a general microscopic theory
of this phenomenon taking into account a multi-level structure of 
tunneling systems in glasses coupled with a quadrupole nuclear moment.
To this end we have developed a simple diagram technique describing the
coherent echo signal. Making use of this technique, one can easily
generalize known results for the echo amplitude from the usual two-level
systems (TLS) to more general multi-level systems.

\section{DIAGRAM REPRESENTATION OF THE ECHO} 

In this section we give a simple diagram representation of the
two-pulse echo signal from an arbitrary multi-level system. We then apply
this technique to a multi-level system formed by a TLS with a
quadrupole nuclear moment. The microscopic details of the
TLS-quadrupole interaction in glasses are discussed further down in 
the next
section. First we start with a simple example of an echo in an ensemble
of TLS. To avoid unnecessary complications we will use a simple
perturbative approach regarding the applied electric field, 
similar to the one considered in Ref.~\cite{GMP}.  However
all results can easily be generalized to strong electric
fields.

\subsection{Echo in an Ensemble of TLS}

The wave function of a TLS in an external ac-electric
field is a linear combination of the wave functions $\varphi_1$ and 
$\varphi_2$ for each level
\begin{equation}
\Psi = C_1\varphi_1+C_2\varphi_2 , \qquad |C_1|^2+|C_2|^2=1 .
\label{eq:e1}
\end{equation} 
Prior to the action of the first electric pulse we have 
$C_1=1$ and $C_2=0$. Then in the electric field the time variation of
the amplitudes $C_1$ and $C_2$ obeys the equations 
\begin{equation}
i\hbar\displaystyle\frac{dC_1}{dt} = E_1C_1 + V(t)C_2, \quad
i\hbar\displaystyle\frac{dC_2}{dt} = E_2C_2 + V(t)C_1 . 
\label{eq:e2}
\end{equation}  
For the off-diagonal transition matrix element $V(t)$ we have a
following expression during the electric pulse  
\begin{equation}
V(t)=V_{1,2}\cos\omega t , \quad 
\mbox{where} \quad V_{1,2}=({\bf F}_{1,2}
\cdot{\bf m})\frac{\Delta_0}{E} . 
\label{eq:e1a}
\end{equation}  
Here ${\bf F}_{1,2}$ is the electric field amplitude of the first or
the second electric pulse, respectively, ${\bf m}$ is 
the dipole moment of the TLS, $\Delta_0$ the tunneling splitting, 
$E=E_2-E_1$ the TLS energy, and $E_1$ and $E_2$ are
the energies of the ground and the excited states of the TLS,
respectively. The electric field frequency is assumed as
$\hbar\omega\approx E$.
To simplify our equations we will put $\hbar=1$.

After the action of the 1st electric pulse $C_2$ acquires in the first
approximation a finite value proportional to amplitude $V_1$ which is
assumed to be small. During the time interval between the first and
second electric pulses ($0<t<\tau_{21}$) we have  
\begin{equation}
C_2 \propto  (-iV_1)\cdot e^{-iE_2t}, \quad
C_1 \approx  1\cdot e^{-iE_1t} - O(V_1^2) .
\label{eq:e3}
\end{equation}  
In Eq.~(\ref{eq:e3}) for $C_2$ we have omitted the resonance factor
\begin{equation}
\beta_{21}(\tau_1)=\frac{1}{z}\sin\frac{z\tau_1}{2}
\exp\left(-iE_2\tau_1+i\frac{z\tau_1}{2} \right) ,
\label{eq:e3a}
\end{equation}   
where $\tau_1$ is a duration of the first pulse, and
$z=E-\omega$ is a detuning from the resonance. For a transition
between the second and the first level (under the action of the
second pulse, see below) the corresponding resonance factor reads
\begin{equation}
\beta_{12}(\tau_2)=-\frac{1}{z}\sin\frac{z\tau_2}{2}
\exp\left(-iE_1\tau_2-i\frac{z\tau_2}{2} \right) .
\label{eq:e3b}
\end{equation}

These resonance factors are very important in describing the {\it form}
of the echo envelope~\cite{GMP2}. But in the present paper we are
mainly interested in the echo {\it amplitude} as a function of the
time delay $\tau_{21}$ between the two pulses. To make our equations
as simple as possible, we assume that $\tau_{21}$ is much larger
than the duration of any pumping pulse, $\tau_i$. As we will show
below, if the electric pulses $\tau_i$ are sufficiently short (the
exact criterion for a TLS-quadrupole multi-level system will be given
in the last section) such resonant factors are not important for our
purposes and we will not take them into account in the following
analysis (though they can easily be included if necessary).

Let us now discuss what happens just after the action of the second
electric pulse $V_2\cos\omega t$. Using Eq.~(\ref{eq:e2}) we
calculate the corresponding variations of $C_1$ and $C_2$:
\begin{equation}
\begin{array}{rcl}
\delta C_1 & \propto & (-iV_2)\cdot C_2(\tau_{21})\propto
(-iV_2)\cdot(-iV_1)e^{-iE_2\tau_{21}} , \\[5pt]
\delta C_2 & \propto & (-iV_2)\cdot C_1(\tau_{21})\propto (-iV_2)\cdot
e^{-iE_1\tau_{21}} , 
\end{array}
\label{eq:e5}
\end{equation}  
(here for the reasons mentioned above we have omitted the resonance
factors $\beta_{12}(\tau_2)$ and 
$\beta_{21}(\tau_2)$ in $\delta C_1$ and $\delta C_2$, respectively).
Finally, at some moment $t$ after the 2nd pulse we have 
\begin{equation}
\delta C_1  \propto  -V_1V_2e^{-iE_2\tau_{21}}\cdot 
e^{-iE_1(t-\tau_{21})}, \quad 
\delta C_2  \propto  -iV_2e^{-iE_1\tau_{21}}\cdot 
e^{-iE_2(t-\tau_{21})} . 
\label{eq:e6}
\end{equation}  

The amplitude of the two-pulse echo from one TLS is determined by the
average value of the product $\delta C_1\delta C_2^*$ ({\em i.e.} by the
off-diagonal density matrix element) which is proportional to 
\begin{equation}
p_{\rm echo} \propto \delta C_1\delta C_2^*\propto 
-iV_1V_2^2 e^{i(E_1-E_2)\tau_{21}} 
\cdot e^{i( E_2-E_1)(t-\tau_{21})} .
\label{eq:e7}
\end{equation}  
Summing over all resonant TLS we have for the two-pulse echo
amplitude 
\begin{equation}
P_{\rm echo} \propto -i\sum_{\rm TLS} V_1V_2^2e^{iE(t-2\tau_{21})} .
\label{eq:e8}
\end{equation}  
For $t=2\tau_{21}$ the amplitude of the echo has a sharp maximum,  
$P_{\rm echo}\propto -iV_1V_2^2\, N$, where $N$ is the number
of resonant TLS (i.e. TLS with $\hbar\omega\approx E)$. 

Let us now give a simple diagram representation of this result. It will
allow us to generalize our approach to the more complicated case of an
echo in a multi-level system formed, for example, by a tunneling 
particle with
a quadrupole nuclear moment. For simplicity, we assume that the electric
field amplitudes ${\bf F}_{1,2}$ of the two electric pulses
are parallel to each other, i.e. ${\bf F}_i=F_i{\bf e}$, where ${\bf
e}$ is a unit polarization vector. Then we have  the following expression
for the induced dipole moment echo amplitude from a single TLS
\begin{equation}
\begin{array}{ccc}
p_{\rm echo}\propto (-iF_1)\alpha_{21}\cdot e^{-iE_2\tau_{21}}\cdot 
(-iF_2)\alpha_{12}\cdot e^{-iE_1(t-\tau_{21})}
\cdot\left[e^{-iE_1\tau_{21}}\cdot (-iF_2)\alpha_{21}\cdot 
e^{-iE_2(t-\tau_{21})} \right]^*\cdot \alpha_{21}=&&\\[10pt]
= -iF_1F_2^2|\alpha_{12}|^2 
|\alpha_{21}|^2e^{i(E_2-E_1)(t-\tau_{21})-i(E_2-E_1)\tau_{21}}=
-iF_1F_2^2|\alpha_{12}|^2|\alpha_{21}|^2e^{iE(t-2\tau_{21})} .
\end{array}
\label{eq:e9}
\end{equation}  
Here $\alpha_{ij}$ are the off-diagonal dipole transition matrix
elements projections on the direction of the electric field. In
the considered case they are real quantities and given by the usual
expression 
\begin{equation}
\alpha_{12} =  \alpha_{21} = ({\bf e}\cdot{\bf
m})\frac{\Delta_0}{E}. 
\label{eq:e10}
\end{equation}  
Generally they obey the relations $\alpha_{12} =
\alpha^*_{21}$.

\begin{figure}
\centerline{\includegraphics[totalheight=3.8cm,keepaspectratio]{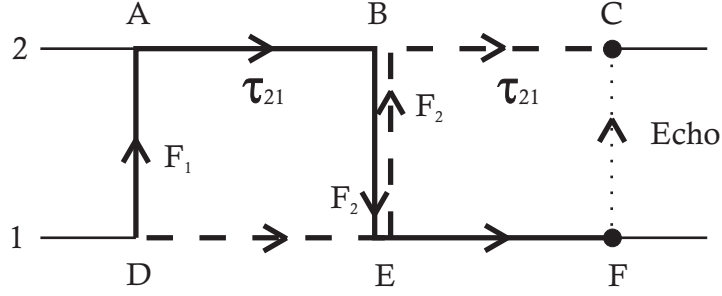}}
\caption{A diagram for a two-pulse echo from a TLS.}  
\label{fig:1}
\end{figure}

Eq.~(\ref{eq:e9}) is similar to Eq.~(\ref{eq:e7}) and 
can be depicted by the diagram shown in Fig.~\ref{fig:1}.
%%%%%%%%%%%%%%%%%%%%%%%%%%%%%%%%%%%%%%%%%%%%%%%
%%%%%%%%%%%%%%%%%%%%%%%%%%%%%%%%%%%%%%%%%%%%%%%%%%
Each full or dashed vertical line in the diagram  corresponds to a
transition from level $i$ to level $k$ ($i,k = 1,2$) under
the action of the first or second electric pulse. To each vertical 
line we ascribe a factor $(-iF_{1,2})\alpha_{ki}$ (depending of the
pulse). For example, the line $DA$ corresponds to the factor
$(-iF_1)\alpha_{21}$ and the full line $BE$ to the factor
$(-iF_2)\alpha_{12}$. The dashed line $EB$ corresponds to the factor 
$(-iF_2)\alpha_{21}$. Finally the dotted line $FC$ (the echo signal)
corresponds to the factor $\alpha_{21}$.
  
Each horizontal line in this diagram (full or dashed)
corresponds to free TLS dynamics in the time intervals between or after
the pulses. Therefore, an appropriate exponential factor is ascribed
to each such line. The factor $\exp(-iE_2\tau_{21})$ corresponds to the
line $AB$ and the factor $\exp(-iE_1\tau_{21})$ to line $DE$.
The factors $\exp[-iE_2(t-\tau_{21})]$ and $\exp[-iE_1(t-\tau_{21})]$
are represented by the lines $BC$ and $EF$, respectively. In the final
expression all factors 
corresponding to dashed lines should be taken as complex conjugated.  
It is easy to see that all full lines (vertical and horizontal)
contribute to the amplitude $\delta C_1$ and all dashed lines to
$\delta C_2$.

\subsection{Echo in a Multi-Level System}
\label{sec:eqms}

\begin{figure}
\centerline{\includegraphics[totalheight=5cm,keepaspectratio]{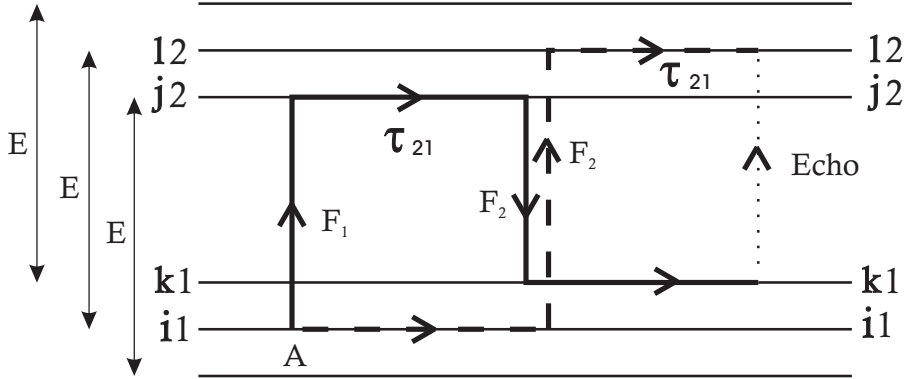}}
\caption{A diagram for the two-pulse echo from TLS-quadrupole
multi-level system.}  
\label{fig:2}
\end{figure}

Using these simple diagram rules we can now easily find the
contributions to the two-pulse echo signal from a multi-level system.
Let us, for example, consider a multi-level system as shown in
Fig.~\ref{fig:2}.  
%%%%%%%%%%%%%%%%%%%%%%%%%%%%%%%%%%%%%%%%%%%%%%
%%%%%%%%%%%%%%%%%%%%%%%%%%%%%%%%%%%%%%%%%%%%%%%%%
It consists of two identical groups of $N$ levels shifted vertically against
each other by some energy $E$ (playing the role of the usual TLS
energy). Inside the groups the positions of the levels are arbitrary, 
i.e. they
are not necessarily equidistant. As we will see in the next section,
such a multi-level system describes a TLS with a quadrupole nuclear
moment.

According to the rules formulated above, the diagram in
Fig.~\ref{fig:2} gives the following partial contributions to the
two-pulse echo signal 
\begin{equation}
\begin{array}{ccc}
\pi_{i1,k1,j2,l2}(t)=\frac{\displaystyle 1}{\sqrt{\displaystyle N}}
(-iF_1)\alpha_{j2,i1}\cdot e^{-iE_{j2}\tau_{21}}\cdot
(-iF_2)\alpha_{k1,j2}\cdot e^{-iE_{k1}(t-\tau_{21})}\times & & \\[10pt]
\times \left[\frac{\displaystyle 1}{\sqrt{\displaystyle N}} 
e^{-iE_{i1}\tau_{21}}\cdot (-iF_2)\alpha_{l2,i1}\cdot
e^{-iE_{l2}(t-\tau_{21})} \right]^*\cdot\alpha_{\rm l2,k1} = & &\\[10pt] 
=-\frac{\displaystyle i}{\displaystyle N}
F_1F_2^2\alpha_{j2,i1}\cdot\alpha_{k1,j2}\cdot\alpha_{l2,i1}^* 
\cdot\alpha_{l2,k1}
\cdot\exp\left[i(E_{l2}-E_{k1})(t-\tau_{21})-
i(E_{j2}-E_{i1})\tau_{21} \right] .
\end{array}
\label{eq:e11}
\end{equation}
The factors $1/\sqrt{N}$ in this formula correspond to the case
when the low-energy group levels are equally populated (with 
a probability $1/N$) and the high-energy group levels are empty. This
is the case for low enough temperatures, $T\ll E$. 
On the other hand, to satisfy the previous conditions the temperature
must be much larger than the width of the energy splitting in the
groups (this width is of the order of the quadrupole splitting
$\Delta E_Q$), i.e. $T\gg\Delta E_Q$. The latter inequality corresponds
to the usual experimental situation. The two conditions are compatible
if $E\gg\Delta E_Q$. In the usual experiments this inequality is obeyed
since the resonance frequency $\hbar\omega\approx E\gg
\Delta E_Q$. The limitation $E\gg T$ is not a crucial.
The final result can be easily generalized to the case 
$T\simeq E$ by including thermal occupation numbers.  

Since the two group of levels are identical and only shifted by 
the TLS energy $E$ we have $E_{l2}=E+E_{l1}$ and $E_{j2}=E+E_{j1}$. Then
Eq.~(\ref{eq:e11}) can be rewritten as
\begin{equation}
\pi_{i1,k1,j2,l2}(t)= -\frac{\displaystyle i}{\displaystyle N}
F_1F_2^2\alpha_{ji}^{(21)}
\alpha_{kj}^{(12)}\alpha_{li}^{*(21)}\alpha_{lk}^{(21)}
\cdot \exp\left[iE(t-2\tau_{21})+i(E_{l1}-E_{k1})(t-\tau_{21})
-i(E_{j1}-E_{i1})\tau_{21}  \right] .
\label{eq:e12}
\end{equation}  
For convenience, we write the indices (1,2) of the level groups as
superscripts. From Eq.~(\ref{eq:e12}) it follows, in the case when
$E\gg\Delta E_Q$, that the echo signal appears at $t=2\tau_{21}$. At this
time, summing over all possible combinations of vertical
transitions ($i,j,k,l$) between different levels and taking into account 
that $\alpha_{ij}^{(12)}=\alpha_{ji}^{*(21)}$, the total 
contribution to the echo signal from one multi-level 
system is~\cite{please} 
\begin{equation}
p_{\rm echo}(2\tau_{21})\propto-\frac{\displaystyle i}{\displaystyle N}
F_1F_2^2 \sum_{i,k}
e^{i(E_i-E_k)\tau_{21}}\left| \sum_j \alpha_{ij}^{(12)}
\alpha_{kj}^{*(12)}e^{iE_j\tau_{21}}\right|^2 .
\label{eq:e12a}
\end{equation}  
This expression differs from the similar one, Eq.~(8) of
Ref.~\cite{WFE}.

\section{TLS INTERACTION WITH NUCLEAR QUADRUPOLES}

In this section we consider a TLS with a  nuclear
quadrupole electric moment in external electric and magnetic
fields. First we derive a Hamiltonian describing the interaction 
and then will apply the perturbation theory approach in respect to this
interaction to describe the echo phenomenon in this system.

\subsection{General Relations}

The Hamiltonian of a nuclear electric quadrupole interacting with a gradient
of an internal electric field has the usual form~\cite{AA}
\begin{equation}
\widehat{\cal H}_{Q}=\widehat{Q}_{ik}\,\varphi_{ik}, \quad 
\varphi_{ik}\equiv \frac{\partial^2\varphi}{\partial r_i\partial r_k} ;
\quad \quad \varphi_{ii}=\Delta\varphi =0 .
\label{eq:14}
\end{equation}  
Here $\varphi$ is an electrostatic potential at the nuclear site
and the traceless tensor 
\begin{equation}
\widehat{Q}_{ik}=\frac{eQ}{6J(2J-1)}\left[\frac{3}{2}(\hat{J}_i\hat{J}_k
+ \hat{J}_k\hat{J}_i)
-\delta_{ik}J(J+1) \right]
\label{eq:15}
\end{equation}  
is the operator of the nuclear quadrupole electric moment, $\hat{{\bf J}}$
is the operator of the nuclear spin.  

The interaction of a nuclear magnetic moment with a magnetic field
$\bf H$ can be written as
\begin{equation}
\widehat{\cal H}_{\rm H}=-\gamma\hbar \hat{{\bf J}}\cdot{\bf H} 
\equiv -\widehat{\bf M}\cdot{\bf H} 
\label{eq:16}
\end{equation}  
where $\gamma$ is the nuclear gyromagnetic ratio and $\widehat{\bf M}$ is 
the nuclear magnetic moment operator. 
Both Hamiltonians (\ref{eq:14}) and (\ref{eq:16}) are Hermitian
$N\times N$ matrices, with $N=2J+1$. 

The total Hamiltonian of a tunneling particle (or a group of particles)
with a quadrupole nuclear moment in an external electric and magnetic
field can be written in as 
\begin{equation}
\widehat{{\cal H}}_{\rm tot} = V(x) - 
{\bf F}\cdot{\bf d}_0\, 
\frac{x}{x_0} + \widehat{Q}_{ik}\,\varphi_{ik}(x) -
\widehat{\bf M}\cdot{\bf H} .
\label{eq:17}
\end{equation}  
Here $x$ is a generalized coordinate of the tunneling particle,
$V(x)$ is a soft atomic double-well potential~\cite{DAP},
${\bf F}$ is the applied electric field, ${\bf d}_0\, x/x_0$ is the
particle electric dipole moment, and $x_0\approx 1\mbox{\AA}$ is of
the order of interatomic distance. The internal electric field
gradient tensor at the site of the nucleus, $\varphi_{ik}(x)$, 
is a function of the generalized particle coordinate $x$.  

Since the relative displacement of a tunneling particle $x/x_0\ll 1$ we
can expand $\varphi_{ik}(x)$ in the Taylor series and limit ourselves
to the linear approximation 
\begin{equation}
\varphi_{ik}(x)=\varphi_{ik}(0) + \varphi_{ik}'(0)\,\frac{x}{x_0} .
\label{eq:19}
\end{equation}   
The second rank tensors $\varphi_{ik}(0)$ and $\varphi_{ik}'(0)$ are
independent of each other and are of the same order of the magnitude,
$\varphi_{ik}'(0)\simeq\varphi_{ik}(0)$. Since the electric field 
gradient is a traceless tensor, $\varphi_{ii}(x)\equiv
0$ for any $x$, the same property holds for the tensors
$\varphi_{ik}(0)$ and $\varphi_{ik}'(0)$, i.e. 
$\varphi_{ii}(0)=\varphi_{ii}'(0)=0$.

As a result the total Hamiltonian (\ref{eq:17}) becomes
\begin{equation}
\widehat{{\cal H}}_{\rm tot} =
V(x) - \widehat{\bf M}\cdot{\bf H} + 
\widehat{Q}_{ik}\,\varphi_{ik}(0) -  
{\bf F}\cdot{\bf d}_0\, \frac{x}{x_0}
+ \widehat{Q}_{ik}\,\varphi_{ik}'(0)\,\frac{x}{x_0} .
\label{eq:21}
\end{equation}  
The first term in this expression is the potential energy of the
tunneling particle. The second and the third terms describe the
interaction of the nuclear magnetic moment with the magnetic field
and of the quadrupole nuclear moment with the {\it average}
internal electric field gradient, respectively. The fourth term 
describes the
interaction of the particle with the external electric field and
finally the last, and for our theory most important term,
accounts for the interaction of the quadrupole nuclear
moment with the particle "orbital" motion in the soft atomic 
potential $V(x)$.
We will see that this last term is responsible for the electric-dipole
echo oscillations in a magnetic field $\bf H$.  

\subsection{TLS Approximation}

To proceed further we will use the usual TLS approximation, keeping in mind
that the particle moves in a nearly symmetric double-well potential
with two minima at $x_1$ and $x_2\approx -x_1$. 
Taking the zero of the potential energy $V(x)$ as the average of
the values $V(x_1)$ and $V(x_2)$, we can write 
\begin{equation}
V(x_1)=\frac{1}{2}\Delta , \qquad V(x_2)=-\frac{1}{2}\Delta ,
\label{eq:22}
\end{equation}    
where $\Delta$ is the energy difference between the two minima (the TLS
asymmetry). Taking the tunneling under the barrier into account,
we can substitute in this approximation
the double-well potential energy $V(x)$ by a $2\times 2$ matrix
\begin{equation}
V(x) \Longrightarrow \frac{1}{2}
\left( 
\begin{array}{cc}
\Delta& -\Delta_0\\
-\Delta_0& -\Delta \\
\end{array}
\right) ,
\label{eq:23}
\end{equation}  
where $\Delta_0$ is the usual tunneling amplitude. 

Similarly way we can write 
\begin{equation}
x \Longrightarrow -|x_1| \left( 
\begin{array}{cc}
1& 0\\
0& -1 \\
\end{array}
\right) \equiv -|x_1|\,\widehat{\sigma}_z ,
\label{eq:24}
\end{equation}
where $\widehat{\sigma}_z$ is the Pauli matrix. As a result 
\begin{equation}
-{\bf F}\cdot{\bf d}_0 \, \frac{x}{x_0} \Longrightarrow
{\bf F}\cdot{\bf m}\, \widehat{\sigma}_z, 
\quad {\bf m} \equiv {\bf d}_0\, \frac{|x_1|}{x_0} , 
\label{eq:25}
\end{equation}
where $\bf m$ is the TLS dipole moment and
\begin{equation}
\widehat{Q}_{ik}\,\varphi_{ik}'(0)\,\frac{x}{x_0}
\Longrightarrow \widehat{\sigma}_z\otimes \widehat{V}_Q, 
\quad \widehat{V}_Q \equiv
-\widehat{Q}_{ik}\,\varphi_{ik}'(0)\, \frac{|x_1|}{x_0} .
\label{eq:26}
\end{equation}

Introducing the notation
\begin{equation}
\widehat{W}_Q \equiv -\widehat{\bf M}\cdot\mbox{\bf H} + 
\widehat{Q}_{ik}\,\varphi_{ik}(0) ,
\label{eq:27}
\end{equation}
we get for the Hamiltonian (\ref{eq:21}) in the TLS approximation
\begin{equation}
\widehat{{\cal H}}_{\rm tot} = \frac{1}{2}
\left( 
\begin{array}{cc}
\Delta& -\Delta_0\\
-\Delta_0& -\Delta \\
\end{array}
\right)
\otimes \widehat{1}_Q + \widehat{1}_{\sigma}\otimes\widehat{W}_Q 
+ {\bf F}\cdot{\bf m}\, \widehat{\sigma}_z \otimes \widehat{1}_Q
+ \widehat{\sigma}_z\otimes\widehat{V}_Q .
\label{eq:28}
\end{equation}
Here $\widehat{1}_Q$ is a $N\times N$ unit matrix in the space of nuclear
spin $\hat{\bf J}$ and $\widehat{1}_{\sigma}$ is a
$2\times 2$ unit matrix in the TLS, $\sigma$ space. The 
symbol $\otimes$ indicates the
direct product of two matrices. As a result $\widehat{{\cal
H}}_{\rm tot}$ is a $2N\times 2N$ Hermitian matrix in a conjoint
space. The last term in this equation describes the interaction of
the nuclear quadrupole with the TLS motion. It will be responsible for the
echo oscillations in a magnetic field. In the following analysis we
will consider this term to be small and treat it by
standard perturbation theory.  

To proceed further, we should diagonalize the first two terms in
Eq.~(\ref{eq:28}). The first term can be diagonalized 
using a standard unitary transformation in TLS space 
\begin{equation}
\widehat{S}_{\sigma}\,
\left( 
\begin{array}{cc}
\Delta& -\Delta_0\\
-\Delta_0& -\Delta \\
\end{array}
\right)
\, \widehat{S}^{-1}_\sigma = 
\left( 
\begin{array}{cc}
E& 0\\
0& -E \\
\end{array}
\right), 
\quad 
\widehat{S}_\sigma=
\left( 
\begin{array}{cc}
\cos\theta& -\sin\theta\\
\sin\theta& \cos\theta \\
\end{array}
\right) ,
\label{eq:29}
\end{equation}  
where $E=\sqrt{\Delta_0^2+\Delta^2}$ and $\tan 2\theta = 
\Delta_0/\Delta$, $\sin 2\theta = \Delta_0/E$. Under this
transformation the $\widehat{\sigma}_z$ matrix in the third and the
fourth terms in Eq.~(\ref{eq:28}) is transformed, as usual, to 
\begin{equation}
\widehat{S}_{\sigma} \widehat{\sigma}_z \widehat{S}^{-1}_\sigma = 
\widehat{S}_{\sigma}\,
\left( 
\begin{array}{cc}
1& 0\\
0& -1 \\
\end{array}
\right)
\, \widehat{S}^{-1}_\sigma = 
\left( 
\begin{array}{cc}
\frac{\strut\displaystyle\Delta}{\strut\displaystyle E}& 
\frac{\strut\displaystyle\Delta_0}{\strut\displaystyle E}\\[10pt]
\frac{\strut\displaystyle\Delta_0}{\strut\displaystyle E} & 
-\frac{\strut\displaystyle\Delta}{\strut\displaystyle E} \\
\end{array}
\right) .
\label{eq:30}
\end{equation}  

In a similar way we can diagonalize the second term in
Eq.~(\ref{eq:28}) and transform the fourth term as follows
\begin{equation}
\widehat{S}_Q\,\widehat{W}_Q\,\widehat{S}_Q^{-1}\equiv
\widehat{\widetilde{W}}_Q, \quad\mbox{and}\quad 
\widehat{S}_Q\,\widehat{V}_Q\,\widehat{S}_Q^{-1}\equiv
\widehat{\widetilde{V}}_Q .
\label{eq:31}
\end{equation}  
Here $\widehat{\widetilde{W}}_Q$ is a diagonal matrix in the nuclear
spin space $N\times N$. It gives the nuclear quadrupole energies in the
average (over the two minima) internal electric field gradient and 
in the external magnetic field ${\bf H}$. The unitary transformation
$\widehat{S}_Q$ can be found in general only by numerical
diagonalization of the matrix $\widehat{W}_Q$. The transformed
matrix $\widehat{\widetilde{V}}_Q$ describes the interaction of the
TLS with the nuclear quadrupole moment.  

Finally as a result of these two independent unitary transformations
the total TLS-quadrupole Hamiltonian reads 
\begin{equation}
\widehat{\widetilde{\cal H}}_{\rm tot} = 
\frac{\displaystyle 1}{\displaystyle \strut 2}
\left( 
\begin{array}{cc}
E& 0\\
0& -E \\
\end{array}
\right)
\otimes \widehat{1}_Q +
\widehat{1}_\sigma\otimes\widehat{\widetilde{W}}_Q 
+\underbrace{ 
\mbox{\bf F}\cdot\mbox{\bf m}\,
\left( 
\begin{array}{cc}
\frac{\strut\displaystyle\Delta}{\strut\displaystyle E}& 
\frac{\strut\displaystyle\Delta_0}{\strut\displaystyle E}\\[10pt]
\frac{\strut\displaystyle\Delta_0}{\strut\displaystyle E} & 
-\frac{\strut\displaystyle\Delta}{\strut\displaystyle E} \\
\end{array}
\right)
\otimes\widehat{1}_Q 
}_{\widehat{\mbox{$f$}}}
+ \underbrace{
\left( 
\begin{array}{cc}
\frac{\strut\displaystyle\Delta}{\strut\displaystyle E}& 
\frac{\strut\displaystyle\Delta_0}{\strut\displaystyle E}\\[10pt]
\frac{\strut\displaystyle\Delta_0}{\strut\displaystyle E} & 
-\frac{\strut\displaystyle\Delta}{\strut\displaystyle E} \\
\end{array}
\right)\otimes \widehat{\widetilde{V}}_Q
}_{\widehat{\mbox{$\cal V$}}} .
\label{eq:32}
\end{equation}  
Here the first two terms are diagonal matrices. Together, they give 
two identical groups of levels (determined by the eigenvalues of
$\widehat{\widetilde{W}}_Q$) shifted from one another by the TLS
energy $E$. For convenience we have introduced the abbrevations 
$\widehat{f}$ and
$\widehat{\cal V}$ for the last two terms. This
notation will be used in the next section.   

\section{DIPOLE TRANSITION MATRIX ELEMENTS}
\label{matrixel}

According to Eq.~(\ref{eq:e12a}) the
electric-dipole echo amplitude is determined by the off-diagonal
dipole transition matrix elements, $\alpha^{(12)}_{ik}$, between the
levels in the TLS-quadrupole multi-level system. In the present 
section, we will calculate, using the Hamiltonian (\ref{eq:32}), these
matrix elements by standard perturbation theory. 
The perturbation will be the last term in Eq.~(\ref{eq:32}),
$\widetilde{\cal V}$. 

\begin{figure}
\centerline{\includegraphics[totalheight=5cm,keepaspectratio]{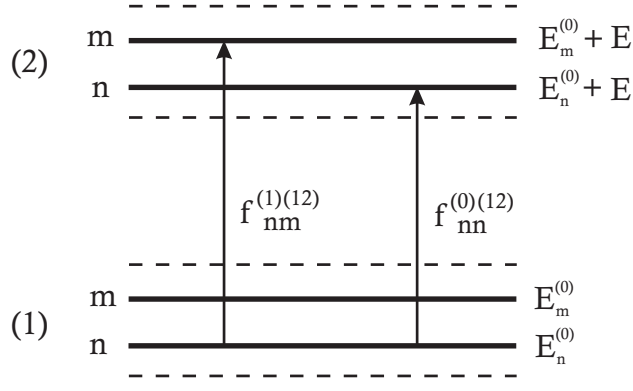}}
\caption{A TLS-quadrupole multi-level system.}  
\label{fig:3a}
\end{figure}

Let us consider the TLS-quadrupole multi-level system shown in
Fig.~\ref{fig:3a}.  
%%%%%%%%%%%%%%%%%%%%%%%%%%%%%%%%%%%%%%%%%%%%%
%%%%%%%%%%%%%%%%%%%%%%%%%%%%%%%%%%%%%%%%%%%%%%%
It consists of two identical groups of levels (1) and (2), shifted by
the TLS energy $E$. We are interested in the off-diagonal transition
matrix elements $f^{(12)}_{nm}$ between these two groups of levels.
In zero order 
($\widehat{{\cal V}}=0$) transitions are induced by the third term in 
Eq.~(\ref{eq:32}). However,
the only non-zero off-diagonal matrix elements are the ones  
for transitions between identical levels in the two
groups 
\begin{equation}
f^{(0)(12)}_{nn}={\bf F}\cdot{\bf m}\frac{\Delta_0}{E}. 
\label{eq:33a}
\end{equation}  
They are all equal and independent of $n$. 

In first order of the perturbation we have also non-zero
off-diagonal matrix elements for transitions between different
levels in the two groups. For $n\ne m$ we have~\cite{LL3} 
\begin{equation}
f^{(1)(12)}_{nm} =
\underbrace{
{\sum_{k\ne n}}\frac{{\cal V}_{nk}^{(11)}f_{km}^{(0)(12)}}
{E_n^{(0)}-E_{k}^{(0)}}
}_{k=m} + 
\underbrace{
{\sum_{k\ne m}}\frac{{\cal V}_{km}^{(22)}f_{nk}^{(0)(12)}}
{E_m^{(0)}-E_{k}^{(0)}}
}_{k=n} = 2 \frac{{\cal V}_{nm}^{(11)}f_{mm}^{(0)(12)}}
{E_n^{(0)}-E_m^{(0)}} .  
\label{eq:34a}
\end{equation}  
Here we have used the property
\begin{equation}
{\cal V}_{nm}^{(11)}=-{\cal V}_{nm}^{(22)} = 
-\frac{\Delta}{E}\left(\widetilde{V}_Q \right)_{nm} ,
\label{eq:35a}
\end{equation}  
which follows from the last term in Eq.~(\ref{eq:32}).

The energy denominators in Eq.~(\ref{eq:34a}) correspond to the
energy difference in one group of levels and are of the order of the
small quadrupole energies $\Delta E_Q$.
In Eq.~(\ref{eq:34a}) we neglected the contributions of the 
off-diagonal matrix
elements ${\cal V}_{nm}^{(12)}$ and the diagonal matrix elements
$f_{nn}^{(0)(11)}$, $f_{nn}^{(0)(22)}$. This is justified since they
have much larger energy denominators, equal to the
distance between two levels from two different groups. This distance
is of the order of the TLS energy $E$ which is much larger than
$\Delta E_Q$. Therefore, this contribution is small due to the small
parameter $\Delta E_Q/E\ll 1$.

Taking into account that for real matrix 
$\left(\widetilde{V}_Q \right)_{nm} = \left(\widetilde{V}_Q \right)_{mn}$, 
we get from Eq.~(\ref{eq:34a}) an important relation for the
off-diagonal matrix elements with $n\ne m$ 
\begin{equation}
f^{(1)(12)}_{nm} = - f^{(1)(12)}_{mn} .
\label{eq:36a}
\end{equation}   
It is equivalent to the important property that
$\alpha^{(12)}_{ij}=-\alpha^{(12)}_{ji}$ for $i\ne j$.  

In the second order in the perturbation we have also corrections to
the off-diagonal transitions matrix elements $f^{(12)}_{nn}$ for
transitions between equal levels in the two groups. Using 
second order perturbation theory~\cite{LL3}
we get after rather cumbersome calculations 
\begin{equation}
f^{(12)}_{nn} = f^{(0)(12)}_{nn}
\left[1-2\sum_{m\ne n}\left(\frac{{\cal V}^{(11)}_{nm}}
{E_n^{(0)}-E_m^{(0)}} \right)^2 \right] .
\label{eq:yh5}
\end{equation}   
In this approximation the matrix elements
$f^{(12)}_{nn}$ differ for different $n$ and are smaller than
the unperturbed values (\ref{eq:33a}).

\section{ECHO OSCILLATIONS IN A MAGNETIC FIELD}

In this section, using the above results, 
we will investigate the echo amplitude as a function of the time delay
$\tau_{21}$ between the two electric pulses. This dependence
straightforwardly give rise to the echo oscillations in the external magnetic
field. For ease of understanding, let us start from the simplest case of
a four-level system. This corresponds to a nuclear 
spin $J=1/2$ in a magnetic field. Such nuclei have zero quadrupole
moment and, therefore, do not interact with a TLS. Nevertheless,
formally we can consider this case to illustrate qualitatively the
main features of our theory. We show that these features will be
conserved in the more realistic case of $J>1/2$. 

\subsection{Four-Level System}

\begin{figure}
\centerline{\includegraphics[totalheight=3.5cm,keepaspectratio]{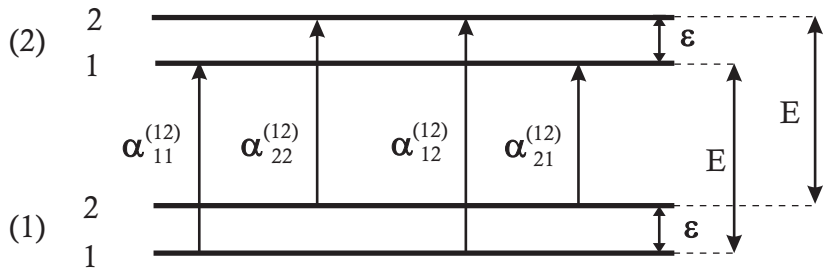}}
\caption{A four-level system for nuclear spin $J=1/2$.}  
\label{fig:4a}
\end{figure}

In the case of $J=1/2$ we have the four-level system shown in
Fig.~\ref{fig:4a}.
%%%%%%%%%%%%%%%%%%%%%%%%%%%%%%%%%%%%%%%
%%%%%%%%%%%%%%%%%%%%%%%%%%%%%%%%%%%%%%%%%%
Introducing the following notations for the off-diagonal matrix elements
$\alpha_{ik}^{(12)}$ between two group of levels (1) and (2)
\begin{equation}
\alpha_{11}^{(12)} = \alpha_{22}^{(12)} =  a_0(1-2b^2) \equiv a,
\quad \alpha_{12}^{(12)} = -\alpha_{21}^{(12)} = 2a_0b,
\label{eq:tv4}
\end{equation}
where
\begin{equation} 
a_0\equiv {\bf e}\cdot{\bf m}\,\frac{\Delta_0}{E} , 
\quad b \equiv \frac{\Delta}{E} 
\frac{\left(\widetilde{V}_Q \right)_{12}}{\varepsilon} ,
\label{eq:37a}
\end{equation}  
and taking into account that in the framework of the perturbation
theory $b\ll 1$, we get from Eq.~(\ref{eq:e12a}) the
following expression for the echo-amplitude from a four-level system 
\begin{equation}
P_{\rm echo} = C\,\left[\left|a^2+4a_0^2b^2e^{i\varepsilon\tau_{21}} 
\right|^2  + 4a_0^4b^2\cos\varepsilon\tau_{21}
\left|1-e^{i\varepsilon\tau_{21}} \right|^2 \right] 
= C\, a_0^4\left(1 - 32b^2\sin^4
\frac{\displaystyle \varepsilon\tau_{21}}{\displaystyle 2}\right)
+ O(b^4) .
\label{eq:38a}
\end{equation}  
Here the coefficient $C$ includes all those factors which are for
the time being irrelevant. 

From Eq.~(\ref{eq:38a}) we can see that the echo amplitude oscillates
as a function of the time delay between the two pulses, $\tau_{21}$. The
frequency of these oscillations is determined by the interlevel splitting
$\varepsilon$ between the levels 1 and 2. Since this splitting is a
function of applied magnetic field the echo will experience
oscillations when magnetic field changes.  

The average level of these oscillations is given by
\begin{equation}
\overline{P}_{\rm echo} = C\, a_0^4(1-12b^2) .
\label{eq:39a}
\end{equation}  
For sufficiently high magnetic fields, the interlevel splitting
$\varepsilon \propto H$ and therefore, according to Eq.~(\ref{eq:37a}), 
$b\propto 1/H \to 0$ for $H\to\infty$. This means that with increasing
magnetic field the average echo amplitude increases and
finally approaches the value of the TLS echo amplitude, $Ca_0^4$ 
without a nuclear quadrupole moment. It looks as if a sufficiently
high magnetic field switched off the TLS-quadrupole interaction.
According to Eq.~(\ref{eq:38a}) we have the same value of the echo
amplitude $Ca_0^4$ in the limit $\tau_{21}\to 0$ but in a final
magnetic field.  

\subsection{TLS-Quadrupole Multi-Level System}

In this section we will generalize our results to the case of an arbitrary
TLS-quadrupole multi-level system. Using a notation similar to
Eqs.~(\ref{eq:tv4}) and (\ref{eq:37a}) 
\begin{equation}
\alpha_{nn}^{(12)}=a_0\left(1-2\sum_{m\ne n}b^2_{nm}\right) ,
\quad \alpha_{nm}^{(12)} = -\alpha_{mn}^{(12)} = 2a_0b_{nm}
\label{eq:iko8}
\end{equation}
where
\begin{equation}
b_{nm}  = \frac{\Delta}{E}\, 
\frac{\left(\widetilde{V}_Q \right)_{nm}}{\varepsilon_{mn}} ,
\label{eq:40a}
\end{equation}  
and $\varepsilon_{mn}=E_m^{(0)}-E_n^{(0)}$ is the distance between
the quadrupole levels (inside one level group!), we get for the echo
amplitude 
\begin{equation}
P_{\rm echo} = C\, a_0^4\left( 1- \frac{64}{N} 
\sum_{n,m>n} b_{nm}^2\sin^4
\frac{\varepsilon_{nm}\tau_{21}}{2}\right) .
\label{eq:41a}
\end{equation}  
Here $N=2J+1$ and the factor C, as before, includes all irrelevant details. 
The average (over $\tau_{21}$) of the echo amplitude is given by  
\begin{equation}
\overline{P}_{\rm echo} = C\,a_0^4\left(1 - \frac{24}{N} 
\sum_{n,m>n}b^2_{nm}\right) .
\label{eq:42a}
\end{equation}  

The quadrupole level splittings $\varepsilon_{mn}$ are functions of
the external magnetic field ${\bf H}$. In the absence of any
level degeneracy, in small magnetic fields when the Zeeman energy $\mu H\ll
\Delta E_Q$, these splittings acquire a small quadratic corrections
of the order of $(\mu H)^2/\Delta E_Q$. Therefore, in this case the
echo amplitude attains, as a function of the applied
magnetic field, oscillations with a frequency of the order of $\mu^2
H\tau_{21}/\hbar\Delta E_Q$. In the case of a level degeneracy at
zero magnetic field, the splittings of the originally degenerate
levels are of the order $\mu H$ and, therefore, the frequency of
the echo oscillations is of the order of $\mu\tau_{21}/\hbar$. A
simple estimate shows that the order of the magnitude of the period
of these oscillations is in a full agreement with the experimental
data~\cite{LESH,LNHE}.

With a further increase of the magnetic field, the Zeeman energy, $\mu H$,
becomes of the order of or bigger than the typical quadrupole energy
splitting $\Delta E_Q$ in  zero magnetic field. In such magnetic fields
the typical level splittings $\varepsilon_{nm}$ are mainly determined
by the Zeeman energy $\mu H$ and in the limit $\mu H\gg \Delta E_Q$
they become linear functions of $H$. Therefore, in this case the
matrix elements $b_{nm}\propto 1/\varepsilon_{nm}\propto 1/H$ are
dying-off with increasing magnetic field. As a result the
average echo amplitude increases, approaching for $H\to\infty$ the
level of the TLS echo amplitude with zero quadrupole moment.

Above we have considered the case of one nucleus interacting with a TLS.
However it can happen in glasses that many nuclei participate in
a tunneling motion. In such a case, since the magnetic 
interaction between the nuclei can be neglected,
we can easily take this into account by
including in Eq.~(\ref{eq:41a}) a summation over the nuclei 
\begin{equation}
P_{\rm echo} = C\, a_0^4\left( 1- \sum_s\frac{64}{N_s} 
\sum_{n,m>n} (b_{nm}^s)^2\sin^4
\frac{\varepsilon_{nm}^s\tau_{21}}{2}\right) ,
\label{eq:43a}
\end{equation}  
where $N_s=2J_s+1$. For distant nuclei the matrix elements $b^s_{nm}$
depend on the distance, $R_s$, between the nucleus and the TLS as
$b^s_{nm}\propto 1/R_s^3$. Therefore, the
main contributions to the sum over nuclei in Eq.~(\ref{eq:43a}) are from
nuclei participating in the tunneling and nearest neighbors.

\subsection{Difference Between Integer And Half-Integer Spins}

Up to now we did not discriminate between the two cases of integer and
half-integer nuclear spin $J$. There is, however, an important
difference between their quadrupole energy level spectra in zero
magnetic field. In the case of an integer spin, $J=1,2,...$, the energy
levels of the quadrupole Hamiltonian $\widehat{W}_Q$ (or
$\widehat{\widetilde{W}}_Q$) in an electric field gradient tensor
$\varphi_{ik}(0)$ of arbitrary symmetry are not degenerate, whereas
in the case of a half-integer spin, $J=3/2, 5/2, ...$, according to
Kramer's theorem all energy levels of the quadrupole Hamiltonian 
$\widehat{W}_Q$ are double degenerate. This degeneracy can only be
lifted by applying a magnetic field. 

To calculate the echo amplitude we have used a perturbation
approach for a non-degenerate case. Strictly speaking this approach
is only valid in a finite magnetic field. Therefore, we now discuss
the special case of a half-integer nuclear spin and 
$H=0$.~\cite{num} In this case the energy
denominators of matrix elements $f_{nm}^{(1)(12)}$ and $b_{nm}$ 
(Eqs.~(\ref{eq:34a}) and (\ref{eq:40a}))
are zero if $n$ and $m$ belong to a pair of Kramer's degenerate levels.   
On the other hand the matrix elements $\left(\widetilde{V}_Q
\right)_{nm}$ are also zero in this case. The reason for this is obvious. 
If the matrix elements were not zero
the perturbation $\widehat{V}_Q$ (or
$\widehat{\widetilde{V}}_Q$) would lift the Kramer's degeneracy of the
nuclear spin levels given by $\widehat{W}_Q$. But this cannot be the case
since 
the interaction of the quadrupole moment with a TLS motion
$\widehat{V}_Q$ belongs to the same symmetry class as $\widehat{W}_Q$.
Therefore, the matrix elements belonging
to a pair of Kramer's degenerated levels $n$ and $n^\prime$ for $H=0$
one has for $H=0$ obey the relations
\begin{equation}
\left(\widehat{\widetilde{V}}_Q^{(0)}\right)_{nn'} =0 \quad
\mbox{for}\quad n\ne n' , \quad 
\mbox{and}
\quad \left(\widehat{\widetilde{V}}_Q^{(0)}\right)_{nn} =
\left(\widehat{\widetilde{V}}_Q^{(0)}\right)_{n'n'} .
\label{eq:44a}
\end{equation}

As a result, we have in our equation for half-integer spin and $H=0$ an
indeterminate form $0/0$ which should be properly 
evaluated. For this we need the limiting behavior of both, 
numerator and the denominator,
as $H\to 0$. It is clear that in a general
the denominator, being the energy splitting between two Kramer's levels
in a magnetic field, vanishes proportional to $H$. The same is true also
for the numerator. The analysis, based on a perturbation theory for 
degenerate levels~\cite{LL3} gives the following expression for 
non-diagonal matrix elements describing the transition between two
Kramer's degenerate levels, $n,n'$, split in an arbitrarily small magnetic
field ${\bf H}$  
\begin{equation}
\left(\widetilde{V}_Q^{(1)}\right)_{nn'} = 
\sum_{m\ne n,n'}\frac{\left({\cal H}_H \right)_{nm} 
\left(\widetilde{V}^{(0)}_Q\right)_{mn'} + 
\left(\widetilde{V}^{(0)}_Q\right)_{nm} 
\left({\cal H}_H \right)_{mn'}} {E_n^{(0)}-E_m^{(0)}} .
\label{eq:45a}
\end{equation}
Here the superscripts $(0)$ and $(1)$ refer to the matrix elements in zero
and non-zero small magnetic fields ${\bf H}$, respectively.
$\left({\cal H}_H \right)_{nm}$ are the matrix elements of the Zeeman
term, Eq.~(\ref{eq:16}). 

We see from Eq.~(\ref{eq:45a}) that the matrix elements
$\left(\widetilde{V}_Q^{(1)}\right)_{nn'}$ are indeed proportional to the
magnetic field $H$ at small fields. Therefore, we come to the
important conclusion that at zero magnetic field the ratio 
$\left(\widetilde{V}_Q^{(1)}\right)_{nn'}/\varepsilon_{nn'}$ has a
{\it finite} value which depends on the orientation of the magnetic
field ${\bf H}$ relative to the principal axes of the internal electric
field gradient tensor $\varphi_{ik}(0)$. Thus, for $H=0$, the
coefficients $b_{nn'}$ for the transitions between Kramer's
degenerate levels are also finite and depend on the
orientation of the applied magnetic field ${\bf H}$. In other words,
we find in this case a non-analytical behavior of the functions
$b_{nn'}({\bf H})$ at ${\bf H}\to 0$.  

This behavior, for half-integer spin, of the matrix elements 
$b_{nn'}$ in a weak magnetic fields
has important consequences. For small magnetic
fields, when the Zeeman energy $\mu H$ is much smaller than the quadrupole
energy $\Delta E_Q$, the distance between the split originally
degenerate 
levels is of the order of $\mu H\ll \Delta E_Q$. On the other hand
the distance between other (non-degenerate) levels is of the order
of $\Delta E_Q$. According to Eq.~(\ref{eq:41a}) a 
contribution to the echo signal from a transition $n\to m$ is
proportional to the oscillating function $(-b^2_{nm}) 
\sin^4(\varepsilon_{nm}\tau_{21}/2)$. This function has a maximum for
$\varepsilon_{nm}=0$. 
Therefore, for small magnetic fields, the contributions of all
quasi-degenerate pairs of levels are {\it in phase}
and proportional to $-\sin^4(\mu H\tau_{21}/2)$. With
increasing magnetic field this function decreases from the
value $0$ to the minimum value -1 for $\mu H\tau_{21}=\pi$. 
In our view, this
explains why in experiment there is often a sharp maximum in the echo
amplitude at zero magnetic field~\cite{LESH,LNHE}. We believe that
this maximum is a contribution from Kramer's degenerate pairs of
levels split by weak magnetic fields and indicates the presence
of nuclei with a half-integer spin $J$ in the glass.    

\section{FINAL REMARKS}

In this final section we address shortly the main approximations we have
made and the limitations of the presented theory. As already
mentioned in the beginning of the paper we have skipped in our
equations all resonance factors like (\ref{eq:e3a}) or
(\ref{eq:e3b}). This is possible when all such factors are
similar 
for all transitions between two group of levels. For this,
the spectral width $\hbar/\tau_i$ of the electric pulses should
be much larger than the typical quadrupole energy splitting $\Delta
E_Q$ which means that the electric pulses should be sufficiently short,
$\Delta E_Q\tau_i/\hbar\ll 1$. Otherwise the important property 
$\alpha^{(12)}_{ij}=-\alpha^{(12)}_{ji}$ for $i\ne j$ which we used
in the paper is not sufficient. The reason is that $i\to
j$ and $j\to i$ transitions correspond to different energy differences (see
Fig.~\ref{fig:4a} for example) and one therefore should multiply
$\alpha^{(12)}_{ij}$ and $\alpha^{(12)}_{ji}$ with different resonance
factors $\beta_{ij}$. In such a case the final expression for the
echo amplitude becomes much more cumbersome. 

The other approximation we made was that we considered the limit of a
rather small electric field amplitudes when the Rabi frequencies
$({\bf F}_{1,2}\cdot{\bf m})\,\Delta_0/\hbar E$ are much smaller than
$\hbar/\tau_i$. In the case of short electric pulses (as was
indicated above) this limitation can be easily overcome and one can
show that Eq.~(\ref{eq:41a}) for the echo amplitude will remain valid
and only the coefficient $C$ changes.  

\section*{ACKNOWLEDGMENTS}
I appreciate many fruitful discussions with S. Hunklinger, C. Enss,
G. Kasper, A. W\"urger, A. Fleischmann, M. Brandt, A. Shumilin. 
I thank Alexander von Humboldt Foundation for their financial support.


\begin{thebibliography}{9}
\bibitem{please} 
Please note that the full and the dashed line should
start from the same level (point $A$ on Fig.~\ref{fig:2}). Otherwise
an additional exponential factor is present. It shows the random initial 
phase difference between the levels and averages to zero.
\bibitem{num} For numerical calculations this problem is
of no importance since for any small but finite value of the
magnetic field our equations are valid.
\end{thebibliography}
\end{document}